\documentclass[10pt,twocolumn,twoside]{IEEEtran}

\usepackage{graphics} 
\usepackage{epsfig} 
\usepackage{amsmath} 
\usepackage{amssymb}  
\usepackage{array}
\usepackage{theorem}
\usepackage[nospace]{cite}
\usepackage[usenames]{color}
\usepackage{subcaption}
\usepackage{todonotes}
\usepackage{algorithm}
\usepackage{algorithmic}
\usepackage{nomencl}
\usepackage{balance}
\usepackage{multirow}
\usepackage{soul}

\theorembodyfont{\rmfamily}
\theoremstyle{definition}\newtheorem{Rem}{Remark}
\theoremstyle{definition}
\theoremstyle{definition}\newtheorem{Cor}{Corollary}
\theoremstyle{definition}\newtheorem{Def}{Definition}
\theoremstyle{definition} \newtheorem{Lem}{Lemma}
\theoremstyle{definition}
\newtheorem{Thm}{Theorem} \theoremstyle{definition}
\newtheorem{Prop}{Proposition} \theoremstyle{definition}

\theoremstyle{definition}
\theoremstyle{definition}

\newcommand{\be}{\begin{equation}}
\newcommand{\ee}{\end{equation}}
\newcommand{\bes}{\begin{equation*}}
\newcommand{\ees}{\end{equation*}}
\newcommand{\bee}{\begin{enumerate}}
\newcommand{\eee}{\end{enumerate}}

\newcommand{\bea}{\begin{eqnarray}}
\newcommand{\eea}{\end{eqnarray}}
\newcommand{\beas}{\begin{eqnarray*}}
\newcommand{\eeas}{\end{eqnarray*}}
\newcommand{\Cov}{\mathrm{Cov}}

\begin{document}


\title{ \LARGE \bf A Potential Game Approach for Information-Maximizing Cooperative Planning of Sensor Networks}

\author{Han-Lim~Choi* and Su-Jin Lee 
\thanks{H.-L. Choi and S.-J. Lee are with the Division of Aerospace Engineering, KAIST, Daejeon, 305-701, Republic of Korea. E-mail: (hanlimc@kaist.ac.kr, sjlee@lics.kaist.ac.kr) }
\thanks{*All correspondence should be forwarded to H.-L. Choi; Mailing address: 291 Daehak-ro, Rm. E4-C327, C-FRIEND Field Robotics Center, KAIST, Yuseong, Daejeon 305-701, Rep. of Korea.; Tel:+82-42-350-3727; E-mail: hanlimc@kaist.ac.kr}
}


\maketitle


\begin{abstract}
This paper presents a potential game approach for distributed cooperative selection of informative sensors, when the goal is to maximize the mutual information between the measurement variables and the quantities of interest. It is proved that a local utility function defined by the conditional mutual information of an agent conditioned on the other agents' sensing decisions leads to a potential game with the global potential being the original mutual information of the cooperative planning problem. The joint strategy fictitious play method is then applied to obtain a distributed solution that provably converges to a pure strategy Nash equilibrium. Two numerical examples on simplified weather forecasting and range-only target tracking verify convergence and performance characteristics of the proposed game-theoretic approach.
\end{abstract}

\section{Introduction}
Mutual information has recently been adopted as a metric of information gain in many contexts (e.g., cooperative sensing for tracking targets~\cite{Grocholsky_PhD02, HoffmanTomlin_TAC10}, weather forecast improvement with mobile sensor networks~\cite{ChoiHow_TCST11, ChoiHow_SJ11, ChoiHow_Auto10}, temperature field described by Gaussian processes~\cite{KrauseGuestrin_JMLR08}, management of deployed sensor networks~\cite{Williams_TSP07, ChoiHowBarton_OPTL12}, adaptive data reduction for simultaneous localization and mapping of mobile robots~\cite{KretzschmarStachiniss_IJRR12}). This popularity results in part from that mutual information is a more natural quantity of information (than entropy) that can be applied to general random entities (i.e., random variables, random processes, random functions, random systems), and in part from that often times sufficient statistics (rather than the actual data) can be used to compute the mutual information. In addition, recent progress on efficient Bayesian inference methods has contributed to effective quantification of mutual information in various contexts.

However, compared to the variety and maturity of the concept of maximizing mutual information for decision making of a network of sensing agents, distributed implementation of such process has not been fully studied and investigated. This may be rather surprising in that the earliest work such as \cite{Grocholsky_PhD02} already addressed distributed/decentralized decision making of this cooperative sensing problem by presenting a local greedy framework in which each agent maximizes the mutual information for its own measurement choice. \cite{HoffmanTomlin_TAC10} extended this local greedy framework and analyzed the optimality gap for such localized decision making to prove that the optimality is monotonically improved as more agents make decisions in a coordinated manner (in their problem setting where submodularity holds). The present author compared the local greedy and a sequential greedy architecture in the context of weather sensor targeting, and pointed out that inter-team information sharing is crucial for performance improvement~\cite{ChoiHow_SJ11}.

This set of previous research provided useful insights and also implementable algorithms; however, the fundamental question of decentralizing the planning process for cooperative sensor networks has not been fully resolved. Compared to the maturity of decentralized/distributed task allocation (e.g., \cite{ChoiBrunetHow_TRO09, Johnson_CDC12, Ger02, Marden_ASME07}), relatively few arguments can be made for sensor network planning. One primary reason for this limit is the underlying correlation structure of the information space in which the cooperative decision is made. For task allocation, for example, one agent's decentralized decision may change the feasibility/fitness of the other agents for that particular task of interest, but there is typically no change in the score structure for other tasks. However, in the cooperative sensing, one agent's decision makes an impact on all other agents' reward structures - for example, if agent 1 selects some task A, then task B that was best for agent 2 is no longer best and task C all of a sudden becomes the best. It is hard to embed this kind of correlation structure in the decentralized decision making. Given this fundamental challenge, this paper tries to take a little bit different approach for distributed decision making for cooperative planning of sensor networks. The authors should point out that the proposed structure is not fully decentralized in the sense that it assumes all the agents know the prior correlation structure correctly and are abe to access to other agents' decisions - thus, it is more of distributed decision making rather than decentralized decision making.

This work presents a game-theoretic approach to address distributed decision making for informative sensor planning. The mutual information is adopted as reward metric to represent the informativeness of sensors. For the global optimization problem of maximizing mutual information between the measurement selection and the quantities of interest, a potential game formulation that defines the local utility functions aligned with the global objective function is proposed. The conditional mutual information of the measurement selection conditioned on the other agents' action enables the potential game formulation, which allows for taking advantage of learning algorithm in the optimization process. The joint strategy fictitious play is particularly considered as an efficient framework to solve the potential game problem. Numerical case studies on idealized weather forecasting and range-only cooperative target tracking using UAVs are presented to demonstrate the feasibility and applicability of the proposed framework. Preliminary work was presented in \cite{Choi_ACC13}, while the present article includes an additional case study that addresses non-Gaussianity in the problem formulation, as well as a refined description of the algorithm.

\section{Preliminaries}
\subsection{Mutual Information} \label{sec:mutual_info}
Mutual information represents the amount of information contained in one random entity ($X$) about the other random entity ($V$).  If the two entities of interest are random vectors, the mutual information can be represented as the difference between the prior and the posterior entropy of $V$ conditioned on $X$:
$$
\mathcal{I}(V; X) = \mathcal{H}(V) - \mathcal{H}(V|X)
$$
where $\mathcal{H}(V) = - \mathbb{E}[ \log (f_{V}(v))]$ represents the entropy of $V$. The mutual information is known to be commutative (or measure-independent~\cite{Cover_Info91}):
$$
\mathcal{I}(V; X) = \mathcal{I}(X; V) = \mathcal{H}(X) - \mathcal{H}(X |V).
$$
When $X$ consists of multiple other random entities, i.e., $X = \{ X_1, X_2, \dots, X_k, \dots, X_n \}$, each of $X_k$ being a random variable and/or a random vector with smaller cardinality, by the chain rule the mutual information can be expressed as:
\bea
\mathcal{I}(V; X) &=& \mathcal{I}(V; X_{k_1}) + \mathcal{I}(V; X_{k_2} |X_{k_1})  \notag \\
&& ~~~ + \dots + \mathcal{I}(V; X_{k_n} | X_{k_1}, \dots, X_{k_{n-1}}) \notag \\
& =& \mathcal{I}(V; X_{k_1}) + \textstyle{\sum_{i=2}^n} \mathcal{I}(V; X_{k_{i}} | X_{k_{1:i-1}}) \label{eq:chain}
\eea
for arbitrary index permutation $\{k_1, \dots, k_n\}$ obtained from $\{1, \dots, n\}$ where $X_{k_{1:i-1}} = \{ X_{k_1}, \dots, X_{k_{i-1}} \}$.

\subsection{Potential Games} \label{sec:potential_game}
Consider a finite game of $n$ players defined by each player's an action set $\mathcal{A}_i$ and a utility function $U_i: \mathcal{A} \rightarrow \mathbb{R}$ with $\mathcal{A} = \times_{i=1}^n \mathcal{A}_i$.  The notation $a_{-i}$ is often used to denote the profile of player actions other than player $i$, i.e., $ a_{-i} = \{ a_1, \dots, a_{i-1}, a_{i+1}, \dots, a_n \}$.

\begin{Def}[Potential Games~\cite{Marden_TAC09}]
A finite $n$-player game with action sets $\{\mathcal{A}_i\}_{i=1}^n$ and utility functions $\{U_i \}_{i=1}^n$ is a \textit{potential game} if, for some scalar potential function $\phi: \mathcal{A} \rightarrow \mathbb{R}$
$$
U_i (a_i', a_{-i}) - U_i (a_i'' , a_{-i} ) = \phi (a_i' , a_{-i} ) - \phi (a_i'', a_{-i})
$$
for every $a_{-i} \in \times_{j \neq i} \mathcal{A}_j$ and for every $a_i', a_i'' \in \mathcal{A}_i$.
\end{Def}

\begin{Def}[Pure Strategy Nash Equilibrium~\cite{Fudenberg_Game91}]
An action profile $a^\star \in \mathcal{A}$ called a \textit{pure strategy Nash equilibrium},  if
$$
U_i (a_i^\star, a_{-i}^\star) = \max_{a_i \in \mathcal{A}_i} U_i (a_i, a_{-i}^\star),\qquad \forall i \in \{1,\dots, n\}.
$$
\end{Def}

\subsubsection{Joint Strategy Fictitious Play~\cite{Marden_TAC09}} \label{sec:jsfp}
As a mechanics to solve a repeated game, this work adopts joint strategy fictitious play (JSFP) developed by \cite{Marden_TAC09}. This section briefs on the basic concept of the JSFP; thus, notations and descriptions are similar to what was written in the original article. JSFP is a sequential decision making process in a \textit{repeated game}, i.e., the same set of games being played over and over again, but with updated set of information at every stage. In JSFP, each player keeps track of the empirical frequencies of the \textit{joint actions} of all other. In contrast to the traditional fictitious play~\cite{Fudenberg_Game91}, the action of player $i$ at stage $t$ is based on the presumption that other players play randomly but jointly according to the joint empirical frequencies.
Let $f_{-i}(a_{-i}'; t) $ be the frequency with which all players but $i$ have selected some joint action profile $a_{-i}'$ up to stage $t-1$:
$$
f_{-i}(a_{-i}'; t) = \frac{1}{t} \sum_{\tau = 0}^{t-1} \mathbb{I} ( a_{-i}(\tau) = a_{-i}' ),
$$
where $\mathbb{I}(\cdot)$ is the indicator function that is unity if the argument is true.
In JSFP, player $i$'s action at stage $t$ is based on the expected utility for action $a_i \in \mathcal{A}_i$, with the joint action model of opponents given by
$$
U_i (a_i ; f_{-i}(t)) = \mathbb{E}_{f_{-i}(t)} \left[ U_i (a_i, a_{-i}) \right]
$$
where the expectation is calculated by the empirical frequency distribution $f_{-i}(t)$ that defines $f_{-i}(a_{-i}; t)$ for all $a_{-i}$. It was pointed out in  \cite{Marden_TAC09} that the predicted utilities $U_i(a_i ; f_{-i}(t))$ for each $a_i \in \mathcal{A}_i$ can be expresses as
 $$
 U_i (a_i, f_{-i}(t)) = \frac{1}{t} \sum_{\tau = 0}^{t-1} U_i (a_i, a_{-i}(\tau))
 $$
 which is the average utility player $i$ would have obtained if he/she had chosen $a_i$ rather than what he/she actually chose at every stage up to time $t-1$ while other players playing the same.  This leads to a recursion~\cite{Marden_TAC09}:
 $$
 U_i(a_i;t+1) = \frac{t}{t+1} U_i (a_i; t) + \frac{1}{t+1} U_i (a_i, a_{-i}(t)),
 $$
 which provides significant computational advantage of JSFP compared to the traditional fictitious play based on empirical frequencies of marginal actions.
 Although convergence of JSFP to a pure Nash equilibrium is not guaranteed even for a potential game, a slight modification of JSFP to include some notion of \textit{inertia} was proven to converge to a pure strategy Nash equilibrium~\cite[Theorem 2.1]{Marden_TAC09}.

\section{Cooperative Sensor Network Planning for Maximum Information} \label{sec:coop_sensing}

\begin{figure*}[t]
\centerline{\includegraphics[width=0.6\textwidth, trim=45 305 45 10,clip]{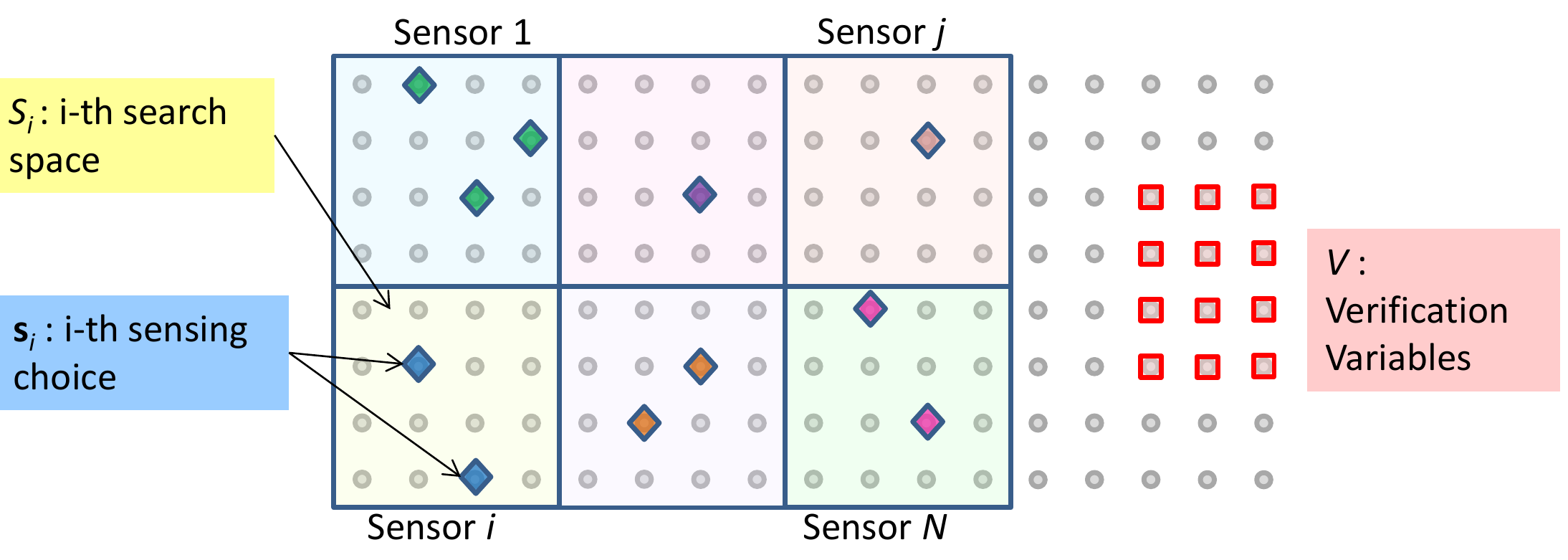}}
  \caption{Concept of Cooperative Planning of Sensor Networks}
  \label{fig:history}
\end{figure*}

Consider a network of $N$ sensing agents\footnote{An agent can be a rather abstract concept depending on the context. Each agent can be a mobile sensor that can move within the specified sensing regions, or a network of sensors among which some sensors will be turned on/off, or possibly a computational unit based on decomposition of the optimization problem.} deployed in a relatively large domain. Each sensing agent is designated to a certain sensing region within which the agent can select points of sensing; agent $i$'s sensing region is denoted as $\mathcal{S}_i$ and is assumed to be finite-dimensional.

Sensing agent $i$ determines a set of sensing points $\mathbf{s}_i = \{s_{i_1}, s_{i_2}, \dots, s_{i_{n_i}} \} \subset \mathcal{S}_i$ with $n_i = |\mathbf{s}_i|$. If an agent takes measurement with some sensing option $s$, the measurement model is given by
$$
Z_s = h_s(X) + W_s
$$
where $W_s$ is some additive sensing noise, and $h_s(\cdot)$ is the observation function that typically is a function of the sensing location $s$. For example, if the agent measures some physical quantity (e.g., temperature, pressure) at the measurement location, $h_s(X) = X_s$ with $X_s$ being the physical entity of interest at $s$; if the agent observes bearing or range to the object of interest from the sensing location $s$, $h_s$ can be expressed as a function of relative geometry of $s$ and the object location.

The goal of \textit{cooperative sensor network planning} is to reduce uncertainty in some random quantity of interest, called \textit{verification variables} and denoted as $V$.  In this work, the mutual information is used to quantify the uncertainty reduction of $V$ by measurement selection by the sensing agents. The mathematical formulation of the cooperative sensing is given by the optimization:
\be %
\mathbf{s}_{1:N}^\star = \arg \max_{\mathbf{s}_{1:N}} ~\mathcal{I}(V; Z_{\mathbf{s}_{1:N}}) \label{eq:coop_sensing} \tag{\textbf{P}}
\ee %
where $ Z_{\mathbf{s}_{1:N}} \triangleq \{ Z_{\mathbf{s}_{1}}, Z_{\mathbf{s}_{2}}, \dots, Z_{\mathbf{s}_N} \}$, which  maximizes mutual information between the verification variables and the measurement chosen by the sensing agents. The verification variables $V$ is some subset of the state variables $X$ but is not necessarily a part of the states associated with some sensing region. It can be state variables associated with a separate region of interest or it can be the whole state variables, depending on the contexts.
The optimization (\ref{eq:coop_sensing}) may be subject to some constraints such as cardinality constraint in each agent's measurement selection, i.e., s $ | \mathbf{s}_i | = n_i. $

If there exists a central agent (or computational unit) that can decide on all the agents sensing decision, (\ref{eq:coop_sensing}) can simply be considered as a combinatorial optimization problem that considers every admissible measurement options $\mathbf{s}_{1:N}$ and computes the mutual information and then chooses the best solution. This process typically requires huge amount of computational resource; the author's previous work provided some techniques such as backward selection based on commutativity of mutual information that can reduce the computational cost of the optimization process~\cite{ChoiHow_TCST11,ChoiHow_SJ11}.

\subsection{Greedy Approximations} \label{sec:dist_opt}
Sometimes more distributed (or decentralized) decision architecture is needed for cooperative sensing problem. One reason is computational complexity. The complexity of (\ref{eq:coop_sensing}) easily grows as the number of agents $N$ increase, number of selected sensing points for each agent $n_i$ increases, and/or the size of sensing region $|\mathcal{S}_i |$ increases. Thus, some level of decomposition and/or approximation is needed to obtain the solution to (\ref{eq:coop_sensing}) with maintaining computational tractability. Another reason is underlying architecture of the network of the sensing agents  where a central unit cannot be assumed due to accessibility of the agent-specific information. The third possible reason is for better robustness to changes and/or uncertainty in the environment.

From these perspectives, there are a couple of approximation techniques that are often adopted: local greedy and sequential greedy decisions.
For the local greedy strategy, agent $i$ simply tries to maximize the uncertainty reduction of $V$ by the measurement itself selects:
\be %
\max  \mathcal{I}(V; Z_{\mathbf{s}_i}), ~\forall i \in \{1, \dots, N\}. \label{eq:lga}
\ee %
This local greedy is known to lead to arbitrary suboptimal result because it totally ignores the coupling and correlation between one agent's measurement selection an the other agents' selections. For this reason, the sequential greedy strategy is frequently adopted, in which agents assume some fixed order of information flow, and an agent optimizes its decision based on the preceding agents' decisions.
\be %
\max \mathcal{I}(V; Z_{\mathbf{s}_{k_i}} | Z_{\mathbf{s}_{k_1}}, \dots, Z_{\mathbf{s}_{k_{i-1}}} ), \forall i \label{eq:sga}
\ee %
for some given index permutation $\{k_1, k_2, \dots, k_N\}$. This sequential greedy approach has been popular, because it often provides good enough suboptimal solutions. In particular, if the underlying mutual information satisfied submodularity, the sequential greedy strategy provides some worst-case performance guarantee: $ \textbf{SGA} \geq (1 - 1/e) \textbf{OPT}$ where \textbf{SGA} and \textbf{OPT} denote objective values for the sequential greedy and the optimal solution, respectively~\cite{KrauseGuestrin_JMLR08}. Submodularity, for example, holds when $V = X$. Also, unlike the objective function in the local greedy strategy, for sequential greedy strategy, the global reward is summation of the local reward because of chain rule of mutual information in (\ref{eq:chain}).

Although the sequential greedy strategy provides a relatively good result, it is still subject to some limitations. Choice of index permutation is not systematic, and more importantly it does not fully take advantage of possible information flows. The strategy solves only one problem for each agent, without considering any possibility of improving the solution performance by communicating agents' decisions.  This work aims at a systematic decision update procedure, inspired by the theory of potential games to achieve closer-to-optimal solution quality.

\section{Main Results} \label{sec:main}
\subsection{Cooperative Sensor Network Planning as Potential Game} \label{sec:sensing_game}
From a game-theoretic perspective, each sensing agent is considered as a selfish entity who simply tries to maximize its local utility function, $U_i(Z_{\mathbf{s}_i}, Z_{\mathbf{s}_{-i}})$ where $Z_{\mathbf{s}_{-i}} \triangleq Z_{\mathbf{s}_{1:N}} \setminus Z_{\mathbf{s}_i}$, which as it is represented is a function of agent $i$'s sensing decision and the decision of the rest of agents. In order for this $N$-person game to provide a solution to the cooperative sensing problem in (\ref{eq:coop_sensing}), the local utility function should be chosen/designed to align with the objective function $\mathcal{I}(V; Z_{\mathbf{s}_{1:N}})$. Being ``aligned'' here means that local player's decision always helps on accomplishing the global objective. The concept of potential game described in section \ref{sec:potential_game} provides a systematic description of this alignment of the local and global objectives. In other words, if one can devise a potential game in which the global potential function is identical to the objective function of (\ref{eq:coop_sensing}), a Nash equilibrium of that potential game can provide a good solution to the cooperative sensor network planning problem.

This section presents one such case as the following Lemma.
\begin{Lem} \label{lem:sensing_game}
With local utility function defined as
\be %
U_i (Z_{\mathbf{s}_i}, Z_{\mathbf{s}_{-i}} ) = \mathcal{I}(V; Z_{\mathbf{s}_i} | Z_{\mathbf{s}_{-i}} ), \label{eq:Ui}
\ee %
the distributed procedure leads to a potential game with global potential:
\be %
\phi(Z_{\mathbf{s}_i}, Z_{\mathbf{s}_{-i}} ) = \mathcal{I}(V; Z_{\mathbf{s}_i}, Z_{\mathbf{s}_{-i}}) =  \mathcal{I}(V; Z_{\mathbf{s}_{1:N}}). \label{eq:phi}
\ee %
\begin{proof}
Consider two different possible sensing actions of agent $i$, $Z_{\mathbf{s}_i}'$ and $Z_{\mathbf{s}_i}''$. Then, the difference in the local utility function with other agents action fixed:
\begin{align*}
& U_i (Z_{\mathbf{s}_i}', Z_{\mathbf{s}_{-i}} ) - U_i (Z_{\mathbf{s}_i}'', Z_{\mathbf{s}_{-i}} ) \\ &~= %
\mathcal{I}(V; Z_{\mathbf{s}_i}' | Z_{\mathbf{s}_{-i}} ) - \mathcal{I}(V; Z_{\mathbf{s}_i}'' | Z_{\mathbf{s}_{-i}} ) \\
&~= \left[ \mathcal{I}(V; Z_{\mathbf{s}_{-i}} ) + \mathcal{I}(V; Z_{\mathbf{s}_i}' | Z_{\mathbf{s}_{-i}} ) \right] \\
&~~~~-  \left[ \mathcal{I}(V; Z_{\mathbf{s}_{-i}} ) + \mathcal{I}(V; Z_{\mathbf{s}_i}'' | Z_{\mathbf{s}_{-i}} ) \right] \\
&~= \mathcal{I}(V; Z_{\mathbf{s}_i}', Z_{\mathbf{s}_{-i}}) - \mathcal{I}(V; Z_{\mathbf{s}_i}'', Z_{\mathbf{s}_{-i}})  \\
&~= \phi (Z_{\mathbf{s}_i}', Z_{\mathbf{s}_{-i}}) - \phi (Z_{\mathbf{s}_i}'', Z_{\mathbf{s}_{-i}}),
\end{align*}
which means the local utility in (\ref{eq:Ui}) leads to a potential game aligned with the global potential (\ref{eq:phi}).
\end{proof}
\end{Lem}

\begin{Rem}
Note that once other agents' actions are fixed at $\mathbf{s}^{\star}_i$, the utility of agent $i$ depends only on its own action; thus, the maximized solution $\mathbf{s}^{\star}_i$ will not be dominated by any unilateral variation of agent $i$'s action. Hence, if a solution to the potential game  in Lemma \ref{lem:sensing_game} is found out, that solution is a pure strategy Nash equilibrium.
%
\end{Rem}

\begin{Cor}
The utility function in Lemma \ref{lem:sensing_game} is not only one that is aligned to the potential function $\phi$ in (\ref{eq:phi}). In general, all the local utility functions of the following form constitute a potential game with the same global potential function:
$$
U_i (Z_{\mathbf{s}_i}, Z_{\mathbf{s}_{-i}}) = \mathcal{I}(V; Z_{\mathbf{s}_i}, Z_{\mathbf{s}_{\mathcal{J}(i)}} | Z_{\mathbf{s}_{- \mathcal{J}(i)}} )
$$
where $\mathcal{J}_1(i), \mathcal{J}_2(i) \subset \{1, \dots, N \}$ are some index sets satisfying
\beas
\mathcal{J}_1 \cup \mathcal{J}_2 \cup \{i\} &=& \{1, \dots, N \}, \\
 \mathcal{J}_1 \cap \{i\} = \mathcal{J}_2 \cap \{i\} & = & \mathcal{J}_1 \cap \mathcal{J}_2  = \emptyset
\eeas
\begin{proof}
For $\mathcal{J}_1(i)$ and $\mathcal{J}_2(i)$ satisfying the above conditions,
$$
Z_{\mathbf{s}_{-i}} = Z_{\mathbf{s}_{\mathcal{J}_1(i)}} \cup Z_{\mathbf{s}_{\mathcal{J}_1(i)}}.
$$
The the potential function, i.e., the global objective function of the cooperative sensing, can be written as:
\begin{align*}
\phi(Z_{\mathbf{s}_i}, Z_{\mathbf{s}_{-i}}) &= \mathcal{I}(V; Z_{\mathbf{s}_i}, Z_{\mathbf{s}_{\mathcal{J}_1(i)}},  Z_{\mathbf{s}_{\mathcal{J}_2(i)}} )  \\
&=  \mathcal{I}(V; Z_{\mathbf{s}_{\mathcal{J}_2(i)}})   +  \mathcal{I}(V; Z_{\mathbf{s}_i}, Z_{\mathbf{s}_{\mathcal{J}_1(i)}} | Z_{\mathbf{s}_{\mathcal{J}_2(i)}})
\end{align*}
by the chain rule of the mutual information.  Thus, for two different actions of agent $i$,
$Z_{\mathbf{s}_i}'$ and $Z_{\mathbf{s}_i}''$
\begin{align*}
&U_i(Z_{\mathbf{s}_i}', Z_{\mathbf{s}_{-i}}) - U_i(Z_{\mathbf{s}_i}'', Z_{\mathbf{s}_{-i}}) \\
&~=  \left[\mathcal{I}(V; Z_{\mathbf{s}_i}', Z_{\mathbf{s}_{\mathcal{J}_1(i)}},  Z_{\mathbf{s}_{\mathcal{J}_2(i)}} )  - \mathcal{I}(V; Z_{\mathbf{s}_{\mathcal{J}_2(i)}})  \right] \\
 &~~~~~- \left[ \mathcal{I}(V; Z_{\mathbf{s}_i}'', Z_{\mathbf{s}_{\mathcal{J}_1(i)}},  Z_{\mathbf{s}_{\mathcal{J}_2(i)}} )  - \mathcal{I}(V; Z_{\mathbf{s}_{\mathcal{J}_2(i)}})  \right] \\
&~= \phi(Z_{\mathbf{s}_i}', Z_{\mathbf{s}_{-i}})  - \phi(Z_{\mathbf{s}_i}', Z_{\mathbf{s}_{-i}})
\end{align*}
\end{proof}
\end{Cor}

\subsection{Iterative Algorithm with Joint Strategy Fictitious Play} \label{sec:sensing_jsfp}
Consider a repeated version of the potential game presented in Lemma \ref{lem:sensing_game}. In a repeated game, agent $i$ makes its decision at stage $t$ based on the information about the other agents' action obtained up to stage $t-1$. For this repeated game, other agents' current action is not available to an agent; thus, the utility function for stage $t$ needs only to be defined as its present action and previous history of the other agents' actions.

One possibly implementation of the potential game in Lemma \ref{lem:sensing_game} is a one-step look-back scheme in which agent $i$ presumes that the other agents made reasonable decisions at the previous stage, and thus maximize local utility conditioned on the other agent's last stage's action:
\be %
U_i(Z_{\mathbf{s}_i};t) = \mathcal{I}(Z_{\mathbf{s}_i} | Z_{\mathbf{s}_{-i}}^\star [t-1]). \label{eq:iga}
\ee %
Although looking simple and implementable, this procedure may not converge to a Nash equilibrium, and as will be demonstrated with numerical examples in section \ref{sec:lorenz}, this often leads to agent's cycling between the actions.

As briefly described in section \ref{sec:jsfp}, the joint strategy fictitious play with inertia proposed by \cite{Marden_TAC09} ensures convergence to a pure Nash equilibrium in the repeated game of potential games. Thus, this paper proposes application of JSFP (with inertia) to solve the potential game for the cooperative sensing problem. The algorithm is summarized in Algorithm \ref{alg:jsfp_sensing}. Some key points to be noted are:
\begin{itemize}
\item At every stage, each agent computes utility values for all of its possible actions. In lines \ref{line:si_for1} and \ref{line:si_for2}, function ${\tt nchoosek}(a,b)$ indicates a set of $b$-dimensional vector whose entry is an integer lest than equal to $a$ with all entries being different each other.
\item In line \ref{line:Ui0}, agents' initial actions are initialized by the solution of local greedy strategy that maximizes mutual information by its own measurement. Note that this initialization step simply provides agents some initial idea about other agents' actions, but does not affect the utility values in the later stages.
\item In line \ref{line:Uit}, at stage $t> 0$, utility values for possible actions are calculated recursively.
\item To introduce inertia in the JSFP process, agents choose its best action at the current stage with probability $\bar{\alpha} \in (0,1)$, while keeping the previous action with probability $1-\bar{\alpha}$ (if statement in line \ref{line:ifbeta}).
\item The convergence is checked by the condition in line \ref{line:term_check}, which is based on calculation of two additional types of actions in lines \ref{line:s_sharp} and \ref{line:s_dagger} as well as the actions chosen by the JSFP process.
\end{itemize}

\begin{algorithm}[t] \label{alg:jsfp_sensing}
\caption{{\sc JSFP for Cooperative Sensor Network Planning}($Z_{\mathcal{S}_{1:N}}$, $V$, $\bar{\alpha} \in (0,1)$)}
\begin{algorithmic}[1]
\STATE $t := 0$
\FOR{ $i \in \{1,\dots, N\}$}
    \FOR{ $\mathbf{s}_i \in {\tt nchoosek}(|\mathcal{S}_{i}|,{n_i})$} \label{line:si_for1}
        \STATE $U_i(Z_{\mathbf{s}_i}; t) = \mathcal{I}(V; Z_{\mathbf{s}_i})$  \label{line:Ui0}
    \ENDFOR
    \STATE $ \mathbf{s}^{\dagger}_i[t] = \arg \max_{\mathbf{s}_i} U_i (Z_{\mathbf{s}};t)$
    \STATE $ \mathbf{s}^{\star}_i[t] = \mathbf{s}^{\dagger}_i[t]$
\ENDFOR
\STATE $ {\tt Coverged } = {\tt FALSE} $
\WHILE{ $\neg~ {\tt Converged}$}
\STATE $ t := t+1 $
\FOR{ $i \in \{1,\dots, N\}$}
    \FOR{ $\mathbf{s}_i \in {\tt nchoosek}(|\mathcal{S}_{i}|,{n_i})$}  \label{line:si_for2}
       \STATE $\widetilde{U}_i (Z_{\mathbf{s}_i}; t) = \mathcal{I}(V; Z_{\mathbf{s}_i}|Z_{\mathbf{s}_{-i}^\star[t-1]})$
        \STATE $U_i(Z_{\mathbf{s}_i}; t) =\frac{1}{t} \widetilde{U}_i(Z_{\mathbf{s}_i}; t)+ \frac{t-1}{t} U_i(Z_{\mathbf{s}_i}; t-1) $ \label{line:Uit}
    \ENDFOR
    \STATE $\mathbf{s}_i^\sharp[t] = \arg \max_{\mathbf{s}_i} \widetilde{U}_i (\mathbf{s}_i; t)$ \label{line:s_sharp}
    \STATE $\mathbf{s}_i^\dagger[t] = \arg \max_{\mathbf{s}_i}  U_i (\mathbf{s}_i; t)$ \label{line:s_dagger}
    \IF{${\tt rand}()  < \bar{\alpha}$} \label{line:ifbeta}
    \STATE $ \mathbf{s}^{\star}_i[t] = \mathbf{s}_i^\dagger[t]$
    \ELSE
    \STATE $ \mathbf{s}^{\star}_i[t] = \mathbf{s}^{\star}_i[t-1] $
    \ENDIF
   \ENDFOR
   \IF{ $\forall i, (\mathbf{s}_i^\sharp[t] \equiv \mathbf{s}_i^\star[t-1]) \wedge (\mathbf{s}_i^\sharp[t] \equiv \mathbf{s}_i^\dagger[t-1]) $ } \label{line:term_check}
   \STATE$ {\tt Converged} = {\tt TRUE} $
   \ENDIF
\ENDWHILE
\end{algorithmic}
\label{alg:jsfp_sensing}
\end{algorithm}

\begin{Lem} \label{lem:termination}
Suppose that the termination condition in line \ref{line:term_check} of Algorithm \ref{alg:jsfp_sensing} is satisfied at some finite $t$. Thus, the following three actions of agent $i$ are identical: $\mathbf{s}_i^\sharp[t]$, $\mathbf{s}_i^\dagger[t-1]$, and $\mathbf{s}_i^\star[t-1]$ for all $i \in \{1, \dots, N \}$; denote this identical action as $\bar{\mathbf{s}}_i$. The following holds for $\bar{\mathbf{s}}_i, i\in \{1,\dots, N\}$:
\begin{enumerate}
\item
$\bar{\mathbf{s}}_i$ is a converged solution, in the sense that:
\begin{eqnarray}
\mathbf{s}^\star_i[t+\tau] = \mathbf{s}^\dagger_i[t+\tau] = \mathbf{s}_i^\sharp[t+ \tau] = \bar{\mathbf{s}}_i, ~\forall \tau \geq 0
\end{eqnarray}
for all $i \in \{1, \dots, N \}$, with consistent tie-breaking in all the $\arg \max$ operations involved.
\item This converged solution $\bar{\mathbf{s}}_{1:N} \triangleq  ( \bar{\mathbf{s}}_1, \dots, \bar{\mathbf{s}}_N)$ is a pure strategy Nash equilibrium of the potential game defined by the utilities in (\ref{eq:Ui}). In other words,
$$
\bar{\mathbf{s}}_i  = \arg \max_{\mathbf{s}_i} \mathcal{I} ( V; Z_{\mathbf{s}_i} | Z_{\bar{\mathbf{s}}_{-i}} ),~\forall i \in \{1,\dots, N\}.
$$
\end{enumerate}
\begin{proof}
\paragraph{Proof of Statement 1} The proof is by induction.
First, consider the case $\tau = 0$ to show that $\mathbf{s}_i^\star[t] = \mathbf{s}_i^\dagger[t] = \bar{\mathbf{s}}_i$, where $\mathbf{s}_i^\sharp[t]=\bar{\mathbf{s}}_i$ by assumption.
The termination condition ensures that $\mathbf{s}_i^\sharp[t] = \mathbf{s}_i^\dagger[t-1] = \bar{\mathbf{s}}_i$, each of which maximizes $\widetilde{U}_i(Z_{\mathbf{s}_i}; t)$ and $U_i(Z_{\mathbf{s}_i}; t-1)$.
Since $U_i(Z_{\mathbf{s}_i};t)$ is a weighted average of $\widetilde{U}_i(Z_{\mathbf{s}_i}; t)$ and $U_i(Z_{\mathbf{s}_i};t-1)$ (from line \ref{line:Uit}), the same action $\bar{\mathbf{s}}_i$ maximizes $U_i(Z_{\mathbf{s}_i};t)$ as well. This means that the best response at stage $t$ becomes:
\begin{equation}
\mathbf{s}_i^\dagger[t] \triangleq \arg \max_{\mathbf{s}_i} = \bar{\mathbf{s}}_i. \label{eq:s_dagger}
\end{equation}
The action chosen in stage $t$, $\mathbf{s}_i^\star[t]$ is determined to be either the best response $\mathbf{s}_i^\dagger[t]$ or the previous action $\mathbf{s}_i^\star[t-1]$, depending on the random number drawn. With (\ref{eq:s_dagger}) and the termination condition, we have $\mathbf{s}_i^\dagger[t] = \mathbf{s}^\star[t-1] = \bar{\mathbf{s}}_i$. Hence, regardless of the random number generated,
$
\mathbf{s}_i^\star[t] = \bar{\mathbf{s}}_i, \label{eq:s_star}
$
which proves the statement for $\tau = 0$.

Second, assume that $\mathbf{s}_i^\star[t+k] = \mathbf{s}_i^\dagger[t+k] = \mathbf{s}_i^\sharp[t+k] = \bar{\mathbf{s}}_i$ for all $k \in \{0, \dots, K-1 \} $. Then, at stage $t+K$
\begin{align*}
\widetilde{U}_i(Z_{\mathbf{s}_i}; t+K) &\triangleq \mathcal{I}(V; Z_{\mathbf{s}_i}|Z_{\mathbf{s}_{-i}^\star[t+K-1]}) \\
&= \mathcal{I}(V; Z_{\mathbf{s}_i}|Z_{\bar{\mathbf{s}}_{-i}}) \triangleq
\widetilde{U}_i(Z_{\mathbf{s}_i}; t)
\end{align*}
for all $\mathbf{s}_i$, which leads to
\begin{equation}
\mathbf{s}^\sharp[t+K] = \mathbf{s}^\sharp[t] = \bar{\mathbf{s}}_i. \label{eq:s_sharp_K}
\end{equation}
From (\ref{eq:s_sharp_K}) and the assumption $\mathbf{s}_i^\dagger[t+K-1] = \bar{\mathbf{s}}_i$, the best response at $t+K$ is obtained as
$
\mathbf{s}^\dagger_i[t+K] = \bar{\mathbf{s}}_i,
$
and  the sameness of $\mathbf{s}^\dagger[t+K]$ and $\mathbf{s}^\star[t+K]$ results in
$
\mathbf{s}_i^\star[t+K] = \bar{\mathbf{s}}_i,
$
which completes the induction proof.

\paragraph{Proof of Statement 2} Consider the best response of agent $i$, when the other agents choose $\bar{\mathbf{s}}_j, \forall  j \neq i $.
$$
\mathcal{I}(V; Z_{\mathbf{s}_i} | Z_{\bar{\mathbf{s}}_{-i}}) = \mathcal{I}(V; Z_{\mathbf{s}_i} | Z_{\mathbf{s}^\star_{-i}[t-1]}) = \widetilde{U}(Z_{\mathbf{s}_i}; t).
$$
By the definition of $\mathbf{s}^\sharp[t] $, which is identical to $\bar{\mathbf{s}}_i$,
$$
\bar{\mathbf{s}}_i = \mathbf{s}^\sharp[t] = \arg \max_{\mathbf{s}_i} \mathcal{I}(V; Z_{\mathbf{s}_i} | Z_{\bar{\mathbf{s}}_{-i}}),
$$
which means that $\bar{\mathbf{s}}_{1:N}$ is a pure strategy Nash equilibrium.

\end{proof}
\end{Lem}

Lemma \ref{lem:termination} proves that if the algorithm is terminated by the specified condition, the resulting solution is a pure strategy Nash equilibrium that is invariant with further stages of game. However, this does not mean that this termination condition is ever satisfied (in finite stages). Fortunately, the original work on the JSFP method ensures convergence of the procedure~\cite[Theorem 2.1]{Marden_TAC09}. To adopt this result, the following can be concluded:

\begin{Thm}
Algorithm \ref{alg:jsfp_sensing} with $\bar{\alpha} \in (0,1)$ almost surely converges to a pure strategy Nash equilibrium of the potential game in Lemma \ref{lem:sensing_game}, with consistent tie-breaking in all the $\arg \max $ operations involved in the process.
\begin{proof}
The iterative procedure represented as the while-loop of Algorithm \ref{alg:jsfp_sensing} constitutes a JSFP process with inertia for a potential game, which is proven to converge to a pure strategy Nash equilibrium almost surely in \cite[Theorem 2.1]{Marden_TAC09}. Also, Lemma \ref{lem:termination} ensures validity of the termination condition of the while loop to stop at a Nash equilibrium. Thus, as a whole, Algorithm \ref{alg:jsfp_sensing} produces a Nash equilibrium for the potential game in Lemma \ref{lem:sensing_game}.
\end{proof}
\end{Thm}

\section{Numerical Examples} \label{sec:numerics}
\subsection{Lorenz-95 Sensor Targeting} \label{sec:lorenz}
The proposed JSFP method is demonstrated on a simplified weather sensor targeting example using Lorenz-95 model.
The Lorenz-95 model~\cite{LorenzEmanuel_JAS98} is an
idealized chaos model that captures key aspects of weather dynamics,
such as energy dissipation, advection, and external forcing. As
such, it has been successfully implemented for the initial
verification of numerical weather prediction
algorithms~\cite{LorenzEmanuel_JAS98, Leutbecher_JAS03}. A two-dimensional extension of the original one-dimensional model was developed and adopted in the author's earlier work~\cite{ChoiHowHansen_ACC07, ChoiHow_TCST11}; this two-dimensional extension represents the global weather dynamics of the mid-latitude region of the northern hemisphere.
The dynamics are:
\be %
\begin{split}
 \dot{y}_{ij} = &\left( y_{i+1,j} - y_{i-2,j} \right) y_{i-1,j}
+ \textstyle{\frac{2}{3}} \left( y_{i,j+1} - y_{i,j-2} \right) y_{i,j-1}  \\& ~~~ - y_{ij} + \bar{y},~ (i=1,\ldots,L_{on}, ~j=1,\ldots,L_{at})
\end{split}
\label{eq:lorenz}
\ee %
where $y_{ij}$ denotes a scalar meteorological quantity, such as
vorticity or temperature~\cite{LorenzEmanuel_JAS98}, at the $(i,j)$-th grid
point; $i$ and $j$ are longitudinal and latitudinal grid indices. There are $L_{on}=36$ longitudinal and $L_{at}=9$
latitudinal grid points; $\bar{y} = 8$ is used for this numerical study. The dynamics are
subject to cyclic boundary conditions in the longitudinal direction ($y_{i+L_{on},j} = y_{i-L_{on},j} = y_{i,j}$) and to the
constant advection condition ($y_{i,0} = y_{i,-1} = y_{i,L_{at}+1}
= 4$ in advection terms) in the latitudinal direction, to model
the mid-latitude area as an annulus.

A similar sensor targeting scenario as \cite{ChoiHow_TCST11} is considered. The problem is to choose optimal sensing locations at time $t_S = 0.05$ (equivalent to 6 hrs in wall clock) to reduce the uncertainty in the verification regions at $t_V = 0.55$ (equivalent to 66hrs).  While a routine network of sensors of size 93 is already deployed and takes measurement every 6 hrs, the decision is to choose additional sensing locations for UAV sensor platforms. The routine network is distributed unevenly to represent data void over the oceanic region. Ensemble square root filter\cite{WhitakerHamill_MWR02} with 1024 samples is used to obtain the correlation structure of the search regions and the verification region. This ensemble is forward integrated to obtain the prior ensemble at the verification time; then, the problem can be treated as a static sensor selection selection problem involving the measurement variables at $t_S$ and the verification variables at $t_V$.  The prior data available from this ensemble forecast pre-processing is Monte-Carlo samples for $ X_{\mathcal{S}_{1:N}} \cup V $, or equivalently the covariance matrix for these variables. The backward selection scheme~\cite{ChoiHow_TCST11} that takes advantage of commutativity of mutual information is utilized to calculate the mutual information:
\begin{align*}
&\mathcal{I}(V; Z_{\mathbf{s}}) \\&~= \mathcal{I}(Z_{\mathbf{s}}; V) \\
&~= \mathcal{H}(Z_{\mathbf{s}}) - \mathcal{H}(Z_{\mathbf{s}}|V) \\
&~= \log \det \Cov(Z_{\mathbf{s}}) - \log \det \Cov (Z_{\mathbf{s}}|V) \\
&~ = \log \det \left( \Cov(X_{\mathbf{s}}) + R_{\mathbf{s}} \right)  - \log \det \left (\Cov (X_{\mathbf{s}}|V) + R_{\mathbf{s}} \right)
\end{align*}
where the measurement equation is given by $Z_s = X_s + W_s$ with $W_s \sim \mathcal{N}(0, R_s)$ for all $s \in \mathcal{S}_{1:N}$. For given covariance $\Cov(X_{\mathcal{S}_{1:N}} \cup V )$, the backward scheme first computes the conditional covariance $\Cov (X_{\mathcal{S}_{1:N}}  |V)  $. Once this conditional covariance is computed, then the selection process for each sensing agent is selection of corresponding principal submatrix and calculation of determinants.

The proposed JSFP-based method is tested for three different sensing topologies -- six sensors in a row, six sensors in 2$\times$3 format, and nine sensors in 3$\times$3 format with equal coverage. An oceanic region of size $12 \times 9$ (in longitude $\times$ latitude) is considered as a potential search region, among which agents are assigned their sensing region $\mathcal{S}_i$. The goal of cooperative sensing is to reduce uncertainty over a separate verification region in time $t_V$; the verification region is not a subset of the search region. The number of sensing points each agent selects, $n_i$, is set to be one for all the cases, the main reason being the optimal solution cannot be obtained in tractable time for larger cases.
For comparison, six different strategies are considered:
\begin{enumerate}
\item Optimal: Optimal solution for the cooperative sensing problem in (\ref{eq:coop_sensing}) is computed by explicit enumeration.
\item Local Greedy: Local greedy strategy that maximizes the mutual information of its own measurement calculated; the solution of local greedy gives the initial condition to iterative algorithms (as in (\ref{eq:lga})).
\item Sequential Greedy: With a fixed agent indexing, sequential greedy algorithm in which an agent selects the point that gives largest mutual information conditioned on the preceding agents' decision (as in (\ref{eq:sga})).
\item Iterative Greedy: An iterative process that only depends on the latest game outcome (as in (\ref{eq:iga})).
\item JSFP w/o Inertia: Implementation of Algorithm \ref{alg:jsfp_sensing} with $\bar{\alpha} = 1.0$.
\item JSFP w/ Inertia: Implementation of Algorithm \ref{alg:jsfp_sensing} with $\bar{\alpha} = 0.3$.
 \end{enumerate}

 The resulting histories of objective values in the iterative procedure are depicted in Fig. \ref{fig:history}.
 Although the cases herein might not be diverse enough to make some concrete statements, some trend can be observed in all the cases (and the similar trend was also found in other unreported test cases).
 \begin{itemize}
 \item JSFP converges to a solution that is same as or close to the optimal solution; the convergence time is less than 20 iterations.
 \item JSFP solution is better than sequential greedy.
 \item No significant difference was observed with varying $\bar{\alpha}$ for JSFP.
 \item Iterative greedy cycles.
    \end{itemize}
It should also be noted that this problem setting of targeting sensor networks to improve forecast in a separate region is particularly difficult case to obtain good solutions using distributed approximation strategies. For example, for the problem of reducing the uncertainty in the entire state (rather than some variables in a particular region), it was observed (although not reported herein) that most strategies result in solutions that are very close to the optimum. The implication is that a certain choice of verification variables may significantly change the underlying information structure.

\begin{figure}[t]
\centerline{\includegraphics[width=1\columnwidth, trim=35 20 40 10,clip]{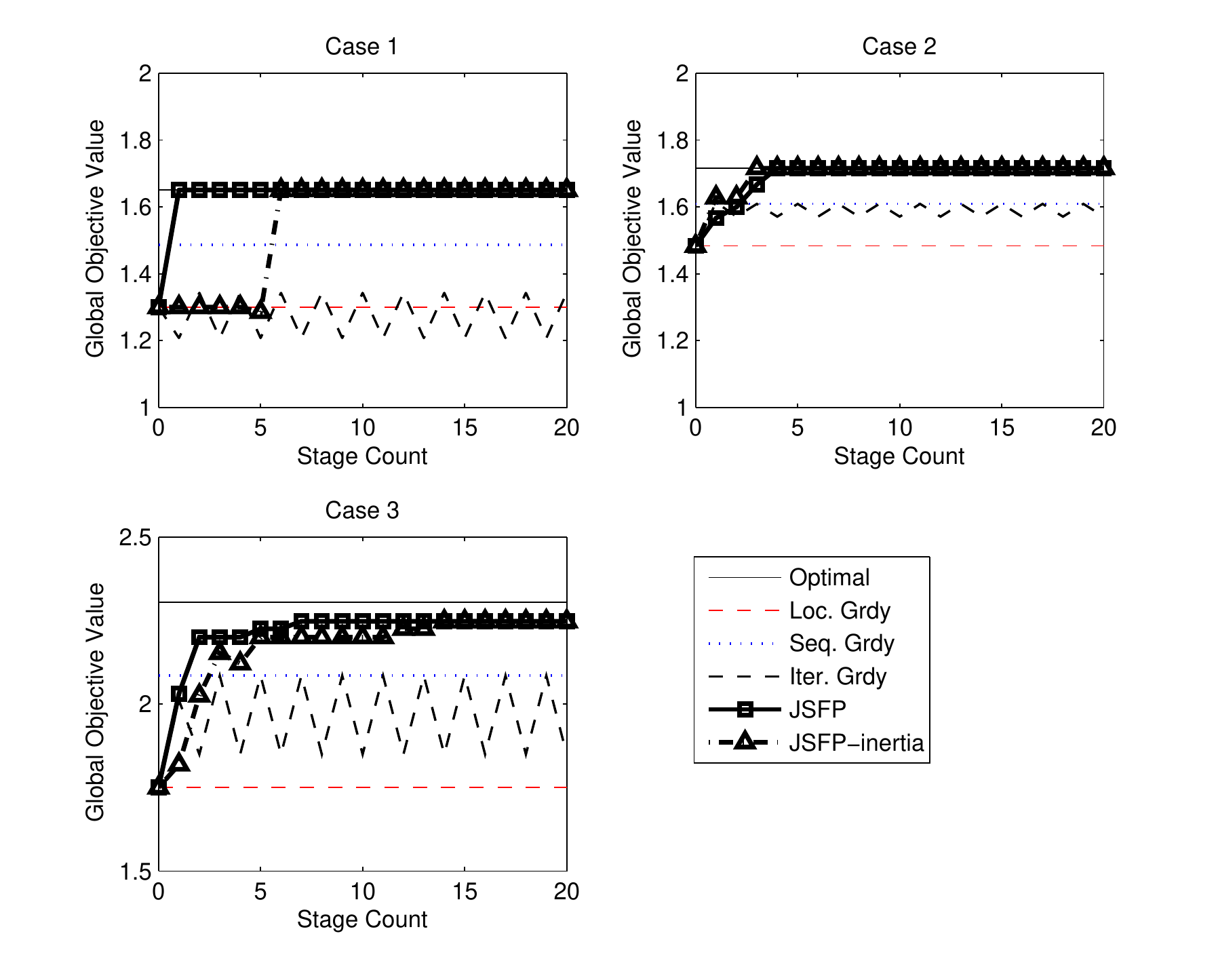}}
  \caption{Histories of Objective Values with Stage Count }
  \label{fig:history}
\end{figure}

\subsection{Range-Only Tracking with UAV Sensors} \label{sec:range}

%

To demonstrate the applicability of the proposed JSFP method in non-Gaussian cases, cooperative localization of a stationary target with range-only sensors equipped on quadrotor UAVs is considered.  The same example was introduced in  \cite{HoffmanTomlin_TAC10}, which presented particle filter-based computation of mutual information (and its local approximation); the present paper takes advantage of this particle filter-based framework in computation of associated mutual information.


 A network of quadrotor sensor platforms try to estimate the location of a stationary target. Each quadrotor carries a particle filter to estimate the target location, while measurements are shared by communication to update the particle filters. The location of quadrotor $i$, denoted as $\mathbf{s}_i$, is represented as the following discrete-time dynamics:
 $$
  \mathbf{s}_i(k+1) = \mathbf{s}_i(k) + a_i(k),
 $$
 where $a_i(k) \in \mathcal{A}_i$ is the control input, which represents movement to one of 12 candidate directions, \{ 0, 45, 90, \dots, 315deg \} with some constant speed. Each UAV takes measurement of the distance between itself and the target with some noise:
 $$
 Z_{\mathbf{s}_i}(k)= || \mathbf{s}_i(k) - \mathbf{r} || + W_i(k)
 $$
 where $\mathbf{r}$ represents the location of the target, and $W_i(k)$  is the additive sensing noise that is not necessarily Gaussian.
To effectively take into account the nonlinearity and non-Gaussianity in the observation process, particle filtering is used to estimate the target state. The particle filter approximates the distribution of the target states with a total of $M$ particles of $(V^{(p)}, \mathbf{w}^{(p)})$, where $V^{(p)}$ and $\mathbf{w}^{(p)}$ represent the states and the importance weight of the $p$th particle. Each agent carries its own particle filter to estimate the target state and utilizes this own filter to make their plans for sensing, while measurements and sensor locations are assumed to be shared by communication among sensors.

At every time step, the sensor network determines best sensing locations to minimize the uncertainty in the target location when the estimate on the target location based on the measurement thus far is available; thus, the problem at time step $k$ can be formulated as the following mutual information maximization:
 \begin{equation}
 \max \mathcal{I}(V; Z_{\mathbf{s}_{1:N}}(k) | \mathcal{Z}_{\mathbf{s}_{1:N}}^{(k-1)}) \label{eq:msn_obj}
 \end{equation}
 where the verification variable $V = \mathbf{r}$ and $\mathcal{Z}_{\mathbf{s}_{1:N}}^{(k-1)} \triangleq \{ Z_{\mathbf{s}_{1:N}}(l) | l \in \{0,\dots, k-1\} \}$. Once the UAV sensors move to the planned sensing locations and take measurements of ranges to the target, these measurements are incorporated into the particle filters of each sensor. The updated particle filter provides prior information for the planning decision at the next time step.

For the global objective in (\ref{eq:msn_obj}), the local utility function for the proposed potential game can be defined as:
$$
U_i(Z_{\mathbf{s}_i}(k), Z_{\mathbf{s}_{-i}}(k)) = 	\mathcal{I}(V; Z_{\mathbf{s}_i}(k) | Z_{\mathbf{s}_{-i}}(k) , \mathcal{Z}_{\mathbf{s}_{1:N}}^{(k-1)}).
$$
Note that this mutual information can alternatively be represented as
\begin{align}
	&\mathcal{I}(V; Z_{\mathbf{s}_i}(k) | Z_{\mathbf{s}_{-i}}(k),\mathcal{Z}_{\mathbf{s}_{1:N}}^{(k-1)} ) \notag \\
	&~= \mathcal{H}(Z_{\mathbf{s}_i}(k)|Z_{\mathbf{s}_{-i}}(k),\mathcal{Z}_{\mathbf{s}_{1:N}}^{(k-1)}) - \mathcal{H}(Z_{\mathbf{s}_i}(k)|V,Z_{\mathbf{s}_{-i}}(k),\mathcal{Z}_{\mathbf{s}_{1:N}}^{(k-1)}) \notag \\
	&~= \mathcal{H}(Z_{\mathbf{s}_i}(k)|Z_{\mathbf{s}_{-i}}(k),\mathcal{Z}_{\mathbf{s}_{1:N}}^{(k-1)}) - \mathcal{H}(Z_{\mathbf{s}_i}(k)|V),
	\label{eq:msn_info}
\end{align}
by the commutativity mutual information and the conditional independence of the measurements conditioned on the target state~\cite{inftheory}. This re-formulation is particularly convenient since the last term $\mathcal{H}(Z_{\mathbf{s}_i}(k)|V)$ represents simply the uncertainty in the measurement noise.

In the particle filtering framework, the computation of the entropy terms in (\ref{eq:msn_info}) can be performed using Monte-Carlo integration and an appropriate numerical quadrature technique presented in , as described in detail in~\cite{HoffmanTomlin_TAC10}.
The  entropy of some measurement $Z_{\mathbf{s}}$ (conditioned on the prior information) can be expressed as
\begin{align*}
\mathcal{H}(Z_{\mathbf{s}}|\mathcal{Z}_{\mathbf{s}_{1:N}}^{(k-1)} ) \approx
-\int_{Z_{\mathbf{s}}}  & \left(\sum_{p=1}^{M} \left( \mathbf{w}^{(p)}p(Z_{\mathbf{s}}|V=V^{(p)}) \right) \right) \\
& \cdot
\log \left(\sum_{p=1}^{M} \left( \mathbf{w}^{(p)}p(Z_{\mathbf{s}}|V=V^{(p)}) \right) \right) dZ_{\mathbf{s}},
\end{align*}
where superscript $(p)$ denotes the particle index. The condition the conditional entropy term can be computed from
\begin{align*}
&\mathcal{H}(Z_{\mathbf{s}_i}(k)|Z_{\mathbf{s}_{-i}}(k),\mathcal{Z}_{\mathbf{s}_{1:N}}^{(k-1)})  \\
&~= \mathcal{H}(Z_{\mathbf{s}_{1:N}}(k)|\mathcal{Z}_{\mathbf{s}_{1:N}}^{(k-1)})  -  \mathcal{H}(Z_{\mathbf{s}_{-i}}(k)|\mathcal{Z}_{\mathbf{s}_{1:N}}^{(k-1)}).
\end{align*}
The conditional entropy of measurement conditioned on the target state can be obtained by
\begin{align*}
\mathcal{H}(Z_{\mathbf{s}_i}|V) \approx &  -\int_{Z_{\mathbf{s}_i}}  \sum_{p=1}^{M} \left\{ \mathbf{w}^{(p)}p(Z_{\mathbf{s}_i}|V=V^{(p)}) \right.\\
&~~\cdot  \left.\log p(Z_{\mathbf{s}_i}|V=V^{(p)}) \right\} dZ_{\mathbf{s}_i}.
\end{align*}
It should be noted that this term does not have any impact on the planning solution unless the distribution of sensing noise varies with the sensing location.

\begin{figure*}[t]
	\centering
	\begin{subfigure}[t]{0.27\textwidth}
		\includegraphics[trim = 10 0 55 45 , clip, width=1\columnwidth]{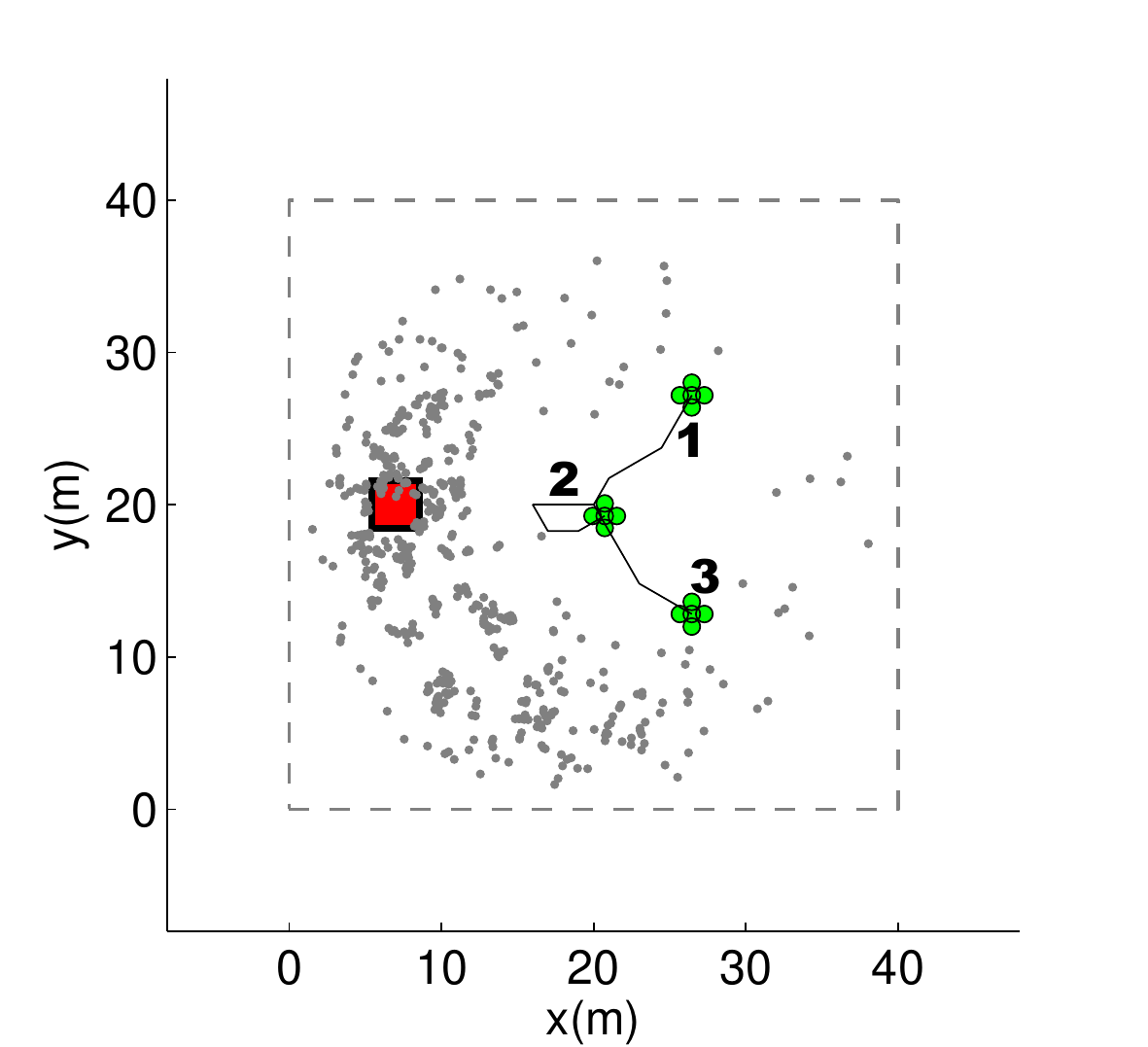}
		\label{fig:msn_simjsfp}
	\end{subfigure}
	\begin{subfigure}[t]{0.27\textwidth}
		\includegraphics[trim = 10 0 55 45 , clip, width=1\columnwidth]{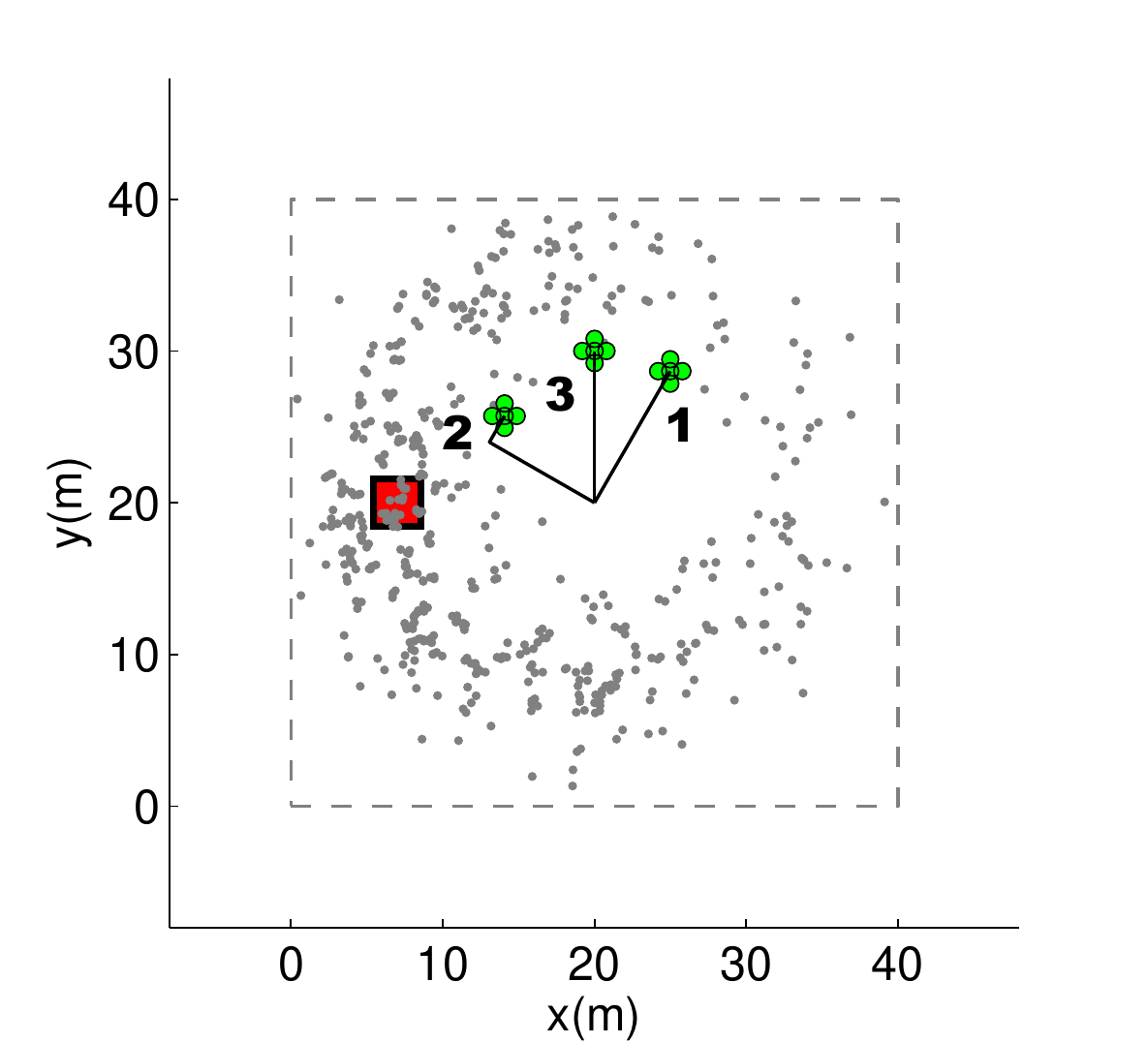}
		\label{fig:msn_simloc}
	\end{subfigure}
	\begin{subfigure}[t]{0.27\textwidth}
		\includegraphics[trim = 10 0 55 45 , clip, width=1\columnwidth]{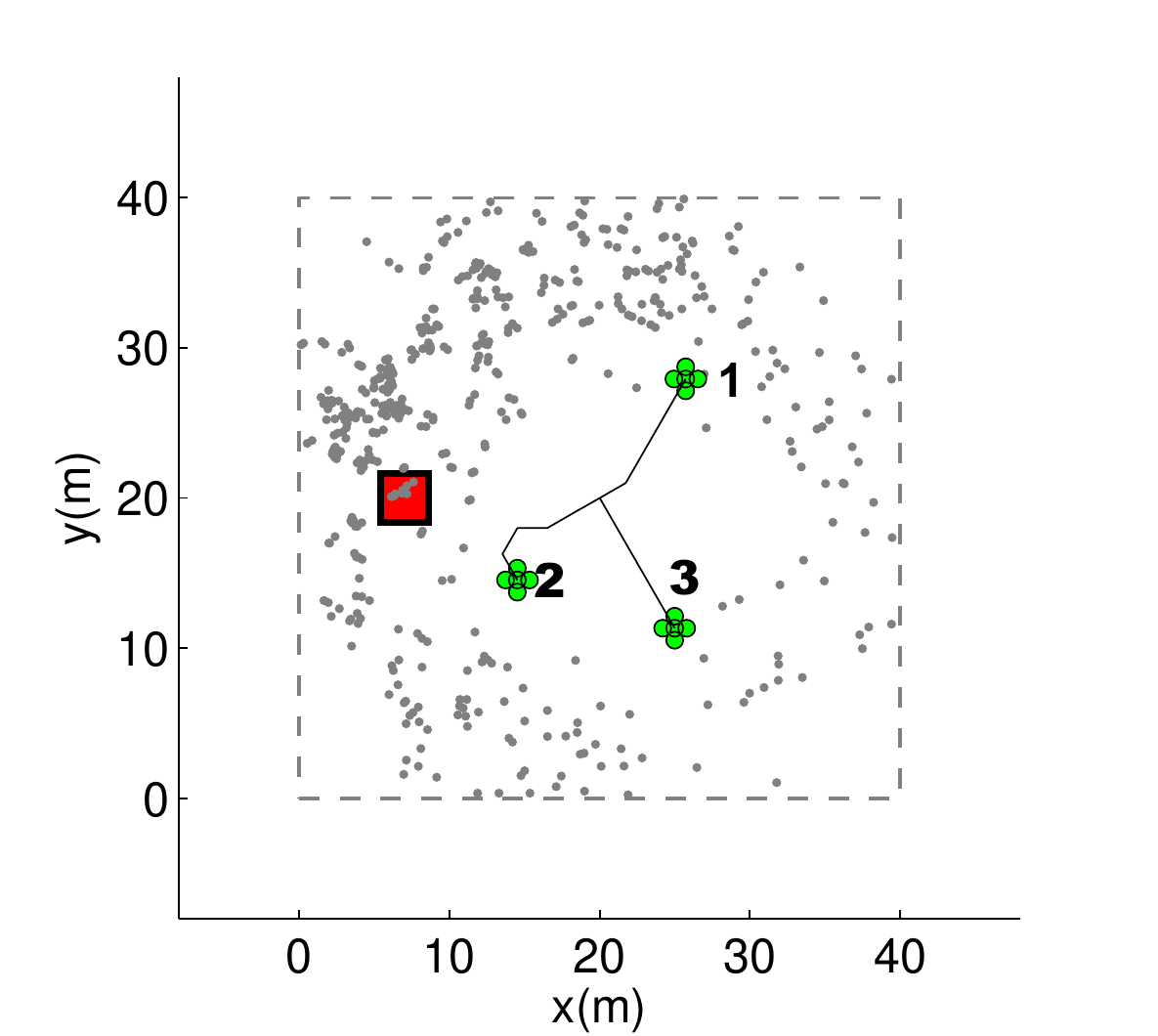}
		\label{fig:msn_simseq}
	\end{subfigure}
	\caption{Illustrative UAV paths for range-only sensing: (left) JSFP, (middle) local greedy, (right) sequential greedy }
	\label{fig:msn_sim}
\end{figure*}

The scenario is to estimate the position of a target  in 40m$\times$40m square region with three UAVs initial positioning at the same location; there is no prior information about the target and particles are uniformly drawn over the whole square region. Fig. \ref{fig:msn_sim} shows the resulting trajectories of the sensor network when using the proposed JSFP method (with $\bar{\alpha} = 0.3$) compared with the local greedy and the sequential greedy strategies. For all three cases, plans are made every time step and the sensors move to the planned positions to take measurements and make decisions for the next time step. The trajectories are overlaid with the particles of the target state carried by the first agent. Observe the difference in the initial direction of UAVs for the three strategies. Since the range-only sensor lacks directional information and all the mobile sensors are initially placed at the same position, the probability distribution of the target location with the measurements taken at the initial time looks like a circle around the UAVs' initial position. Since every direction is equally uncertain, physical insights suggest that the optimal cooperative behavior of the sensor network is to spread out with equal angular distance (i.e., 120degs) to lower down the overall uncertainty level. This type of behavior is observed for the JSFP solution. For the sequential greedy case, it can be seen that two agents who make the first two decisions move to the opposite direction, which is the best way for two-agent problem.  For the local greedy case, agents do not coordinate and move to a similar direction, which is highly suboptimal.

\begin{figure}[t]
	\centering
	\includegraphics[trim = 5 0 20 15, clip, width=1\columnwidth]{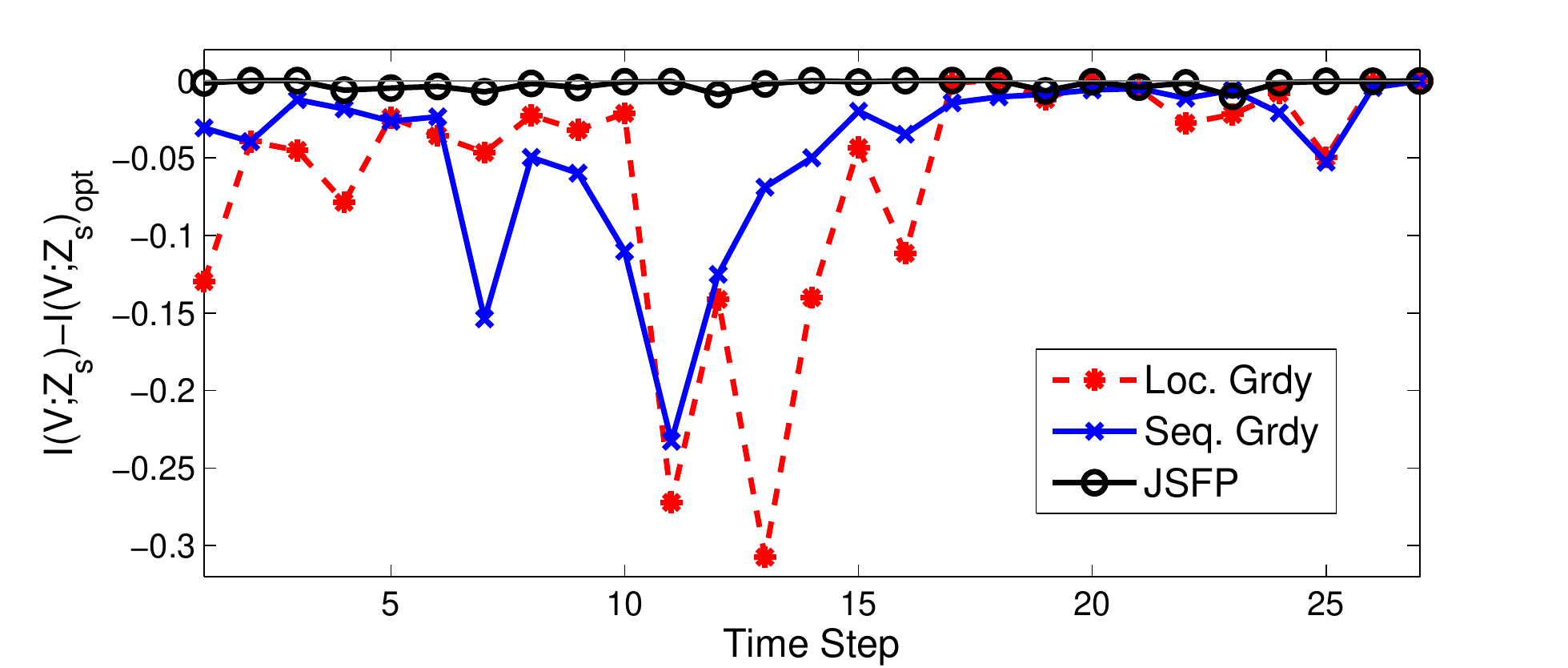}
	\vspace*{-.15in}
	\caption{Optimality gaps for different distributed planning strategies}
	\label{fig:msn_obj}
	\vspace*{-.3in}
\end{figure}

In addition, performance of the JSFP method is further investigated as follows. Each agent moves in a randomly chosen direction and takes measurement; however, every time step the planning problem is posed and the solutions for four different strategies (optimal, JSFP with $\bar{\alpha} = 0.3$, local greedy, and sequential greedy) are computed. Fig. \ref{fig:msn_obj} shows the optimality gaps of the strategies for every time step. Observe that JSFP produces solutions very close to the optimum every cases, while the other greedy methods result in large optimality gap for some cases. The JSFP procedure converges within  5.7 stages on average, and average of 7.1 calculations of mutual information are involved until convergence -- note that the latter is less than three times of the former as an agent computes mutual information values only when the other agents change their decisions. This number of calculations is certainly more expensive than the greedy strategies that do not have any iterative procedures, and is much less than exhaustive search for the optimal solution in which hundreds of calculations are required.


\section{Conclusions}
A potential game-based approach for distributed selection of informative sensors has been presented when the global information reward is defined as the mutual information between the measurement variables and the entities of interest. A local utility function defined that leads to a potential game is proposed, and the joint strategy fictitious play method is then applied to obtain a Nash equilibrium of the potential game. Two numerical examples have demonstrated validity of the proposed method compared to other distributed schemes.

\section*{Acknowledgments}
This work was supported in part by AFOSR grant (FA9550-12-1-0313), and in part by the KI Project via KI for Design of Complex Systems.

\bibliographystyle{IEEEtran}
\bibliography{acc13_game}

\end{document}